\pdfoutput=1
\documentclass[12pt,a4paper]{article}
\usepackage{xcolor}
\usepackage{enumerate}
\usepackage{framed}
\usepackage{mdframed}
\usepackage{amsmath}
\usepackage{amsfonts}
\usepackage{mathrsfs}
\usepackage{amssymb}
\usepackage{graphicx, rotating}
\usepackage{epsfig}
\usepackage{latexsym}
\usepackage{graphicx}
\usepackage{color}
\usepackage{amsmath,bm,amssymb}
\usepackage{cite}
\usepackage{slashed}
\usepackage{hyperref}
\usepackage{epstopdf}
\usepackage[font=footnotesize,labelfont=]{caption}
\AppendGraphicsExtensions{.tif}
\hypersetup{colorlinks=true, citecolor=bluscuro, linkcolor=black, urlcolor=bluscuro}
\definecolor{rossos}{cmyk}{0,1,1,0.55}
\definecolor{bluscuro}{rgb}{0.15, 0.2, .85}
\definecolor{bluchiaro}{cmyk}{1,.3,0.,0.1}

\newcommand{\bea}{\begin{eqnarray}}
\newcommand{\eea}{\end{eqnarray}}
\def\beq{\begin{equation}}
\def\eeq{\end{equation}}
\def\d{{\rm d}}

 at 10truept

\newcommand{\llp}{\left [}
\newcommand{\rrp}{\right ]}
\newcommand{\lp}{\left (}
\newcommand{\rp}{\right )}
\newcommand{\be}{\begin{equation}\begin{aligned}}
\newcommand{\ee}{\end{aligned}\end{equation}}

 at 10truept

\begin{document}
\def\thefootnote{\fnsymbol{footnote}}

\begin{center}
\Large{\textbf{A Note on the Abundance of Primordial Black Holes: \\
Use and Misuse of the Metric Curvature Perturbation}} \\[0.5cm]
\end{center}
\vspace{0.5cm}

\begin{center}

\large{V. De Luca\footnote{valerio.deluca@unige.ch} and A.~Riotto\footnote{antonio.riotto@unige.ch}}
\\[0.5cm]

\small{
\textit{Department of Theoretical Physics and Center for Astroparticle Physics (CAP) \\
24 quai E. Ansermet, CH-1211 Geneva 4, Switzerland}}

\vspace{.2cm}
\vspace{.2cm}
\vspace{.2cm}

\end{center}

\vspace{.7cm}
\hrule \vspace{0.3cm}
\noindent \small{\textbf{Abstract}\\ 
The formation of Primordial Black Holes (PBHs) through the collapse of large fluctuations in the early universe is a rare event. This manifests itself, for instance, through the non-Gaussian tail of the formation probability. To compute such probability and the abundance of PBHs, the curvature perturbation is frequently adopted. In this note we emphasize that its use does not provide the correct PBH formation probability. Through a path-integral approach we show that the exact calculation of the PBH abundance demands the knowledge of multivariate joint probabilities of the curvature perturbation or, equivalently, of all the corresponding connected correlators.}

\vspace{0.3cm}
\noindent
\hrule
\def\thefootnote{\arabic{footnote}}
\setcounter{footnote}{0}


\section{Introduction}
\renewcommand{\theequation}{1.\arabic{equation}}
\setcounter{equation}{0}
\label{intro}
\noindent
The formation of PBHs in the early universe is a rare event (for recent reviews,  see \cite{Green:2020jor,Carr:2021bzv,Franciolini:2021nvv}). 
In the most  standard mechanism  PBHs are originated from the enhancement of the curvature power spectrum below a certain length scale. If the  power spectrum of the curvature perturbation is enhanced during inflation from its  value $\sim 10^{-9}$ at large scales to $\sim 10^{-2}$  on small scales, and after the fluctuations are  transferred to radiation during the reheating process after inflation, PBHs may form from sizeable fluctuations in the radiation density field. This happens  if they are able to overcome the resistance of the radiation pressure (see Ref.~\cite{Sasaki:2018dmp} and references therein for more details). 

Since such PBHs are generated thanks to   large and rare fluctuations beyond a given threshold,  the formation probability, as well as their  mass fraction, are extremely sensitive to tiny changes in the tail of the probability distribution, and therefore to possible 
non-Gaussianities~\cite{Bullock:1996at, Ivanov:1997ia, Yokoyama:1998pt, PinaAvelino:2005rm,  Saito:2008em, Byrnes:2012yx, Bugaev:2013vba, Young:2013oia, Young:2015cyn, Kawasaki:2015ppx, Pattison:2017mbe, Atal:2018neu, Franciolini:2018vbk}.

There is much ambiguity in the literature about which  criterion should be used to identify  the  perturbations  able to form a PBH and  how the corresponding critical threshold is calculated. Nevertheless, it is rather clear by now that the correct quantity to use is the smoothed density contrast $\delta_m$ (to be defined later on) rather than  a metric perturbation such as the comoving curvature perturbation $\zeta$. 

First of all,   on superhorizon scales and by a coordinate transformation,  one can always shift  the comoving curvature perturbation by an arbitrary  constant, making the calculation of the PBH abundance  not physical. This problem is avoided if the  density contrast is adopted, as it depends on spatial derivatives of the curvature perturbation. Second of all, the density  distribution  may be not  already smooth on scales smaller than the scale being considered, and therefore a smoothed density contrast should be used.

While these comments are already present in the literature~\cite{Young:2014ana,Yoo:2018kvb,Young:2019yug,Biagetti:2021eep,Kitajima:2021fpq,Tada:2021zzj}, the confusion seems to persist as, when realized that  a  perturbative treatment of the statistics of rare fluctuations fails~\cite{Vennin:2020kng,Figueroa:2020jkf,Pattison:2021oen,Figueroa:2021zah,Hooshangi:2021ubn,Cai:2021zsp,Achucarro:2021pdh,Hooshangi:2022lao}, the comoving curvature perturbation continues to be adopted in the calculation of the PBH abundance.

The goal of this note is to show that 

\begin{enumerate} 
\item the knowledge of the probability distribution of the comoving curvature perturbation does not suffice to calculate the abundance of PBHs, being the smoothed density contrast the correct variable to use. This in turn requires the knowledge of the probability distribution of the radial derivative $\zeta'$ at a distance $r_m$ (to be defined later on) from the threshold (or peak) and, therefore,  of all the  connected $n$-point correlators of the curvature perturbation. In other words, one needs to know its joint probability
$P[\zeta(\vec x_1),\cdots,\zeta(\vec x_n)]$.

\item As the abundance of PBHs depends on the 
threshold $\delta_c$ of the smoothed density contrast which, in turn,  depends on the (mean) profile of the curvature perturbation $\zeta(r)$ \cite{Germani:2018jgr}, again the knowledge of at least the joint two-point probability
$P[\zeta(\vec x_1),\zeta(\vec x_2)]$ is necessary.  
\end{enumerate} 
We will use a path-integral approach to make these points as transparent as possible. Of course, if the comoving curvature perturbation $\zeta$ is a functional of a Gaussian linear component, whose multivariate joint probability is known, one  should calculate the corresponding  joint probability of the non-linear variable $\zeta$  and the abundance of PBHs passing through the smoothed density contrast and its threshold.

This note is organised as follows. After some preliminaries in Section 2, Section 3 is dedicated to discuss the two main points raised in the introduction. Section 4 offers a comparison of the calculation of the PBH abundance through the smoothed density contrast and the curvature perturbation. Finally our conclusions are given in Section 5.

\section{Some preliminaries}
\renewcommand{\theequation}{2.\arabic{equation}}
\setcounter{equation}{0}
\noindent
As we mentioned in the introduction,  we focus on one of the most likely scenario for the formation of PBHs, the collapse of sizeable overdensities generated during inflation which re-enter the cosmological horizon during the subsequent radiation-dominated era.

On super-horizon scales, one can apply the gradient expansion approach~\cite{Shibata:1999zs} to write down the relation between the non-linear density contrast $\delta(r)$ and the time independent curvature perturbation $\zeta(r)$ as~\cite{Harada:2015yda}
\be
\label{deltaNL}
\delta(r)=-\frac{8}{9}\left(\frac{1}{aH}\right)^2 e^{-5\zeta(r)/2}\nabla^2 e^{\zeta(r)/2},
\ee
in terms of the scale factor $a$ and Hubble rate $H$.
The criterion for which a cosmological perturbation may undertake gravitational collapse and form a PBH depends on the amplitude measured at the peak of the compaction function, defined to be the mass excess compared to the background value in a given radius. It can be expressed on super-horizon scales in terms of the curvature profile as~\cite{Harada:2015yda, Musco:2018rwt}
\be
\mathcal{C} (r) = - \frac{2}{3}\, r\, \zeta' (r) \llp 2 +  r\, \zeta' (r) \rrp,
\ee
where the prime stands for differentiation with respect to $r$.

The compaction function takes its maximum at the comoving length scale $r_m$ satisfying
\be
\mathcal{C}'(r_m) = 0 \qquad  {\rm or}  \qquad  \zeta' (r_m) + r_m \zeta'' (r_m) = 0.
\ee
One can then define consistently the smoothed perturbation amplitude as the volume average of the energy density contrast within the scale $r_m$, measured at the cosmological horizon crossing time $a H r_m = 1$~\cite{Musco:2018rwt}
\be
\delta_m=\frac{3}{\left(r_m e^{\zeta(r_m)}\right)^3}\int_0^{r_m} {\rm d} r\,\delta(r)\left(r e^{\zeta(r)}\right)^2\left(r e^{\zeta(r)}\right)',
\ee
where we have adopted a top-hat window function to account for the treatment of the threshold~\cite{Young:2019osy}.
This represents the main quantity determining the abundance of PBHs, which can be simplified to give
\be
\delta_m = \delta_l - \frac{3}{8} \delta_l^2, \qquad \delta_l = - \frac{4}{3}r_m \zeta' (r_m).
\ee
As one can already appreciate at this stage, only gradients of the comoving curvature perturbation may affect the PBH formation, given that on super-horizon scales one can always perform a coordinate transformation which shifts the $\zeta$, but without affecting the physics. Furthermore, a correspondence between peaks in the curvature perturbation and peaks in the density contrast is guaranteed only at linear level~\cite{Young:2019yug}.
These comments clarify how a well-posed criterion to distinguish which perturbations may collapse to a PBH should be based on the smoothed density contrast and not on the curvature perturbation.

The PBH abundance can then be computed by integrating the probability distribution function of the smoothed density contrast from a threshold value $\delta_c$ on, as
\be
\beta = \int_{\delta_c} P(\delta_m) \d \delta_m.
\ee
However, using the relation shown above, one can use the conservation of the probability to write 
\be
P(\delta_l) \d \delta_l = P(\delta_m) \d \delta_m,
\ee
such that the linear smoothed density contrast is the ultimate key parameter which we have to compute the probability of, with a corresponding threshold given by
\be
\label{threhsold_delta}
\delta_{l,c} = \frac{4}{3} \lp 1- \sqrt{1-\frac{3}{2}\delta_c} \rp.
\ee
It is crucial to note finally that the probability of $\delta_l$ depends on the statistical properties of the curvature perturbation $\zeta$, which are dictated by the dynamics of the formation scenario under consideration.

The curvature perturbation profile is also fundamental to determine the threshold to collapse. In particular, many general-relativity numerical simulations have been dedicated to the computation of the threshold assuming a benchmark parameterization of the curvature profile as (see for instance Refs.~\cite{Musco:2018rwt, Musco:2020jjb})
\be
\label{parametrization}
\zeta (r) = \zeta_0 \exp \llp - \lp \frac{r}{r_m} \rp^{2\gamma} \rrp,
\ee
in terms of the amplitude $\zeta_0$ and scale $r_m$ of the collapsing perturbation.
This corresponds to a energy density profile of~\cite{Young:2019yug}
\be
\delta (r) = \frac{8}{9}\lp \frac{1}{a H} \rp^2 \gamma \lp \frac{r}{r_m} \rp^{2\gamma} \llp (2 \gamma + 1) - \gamma \lp \frac{r}{r_m} \rp^{2\gamma} \lp 2 + \zeta(r) \rp \rrp \frac{\zeta(r)}{r^2 e^{2\zeta(r)}},
\ee
obtained by inserting the assumed parameterization in Eq.~\eqref{deltaNL}.
One can then relate the threshold of the smoothed density contrast to the amplitude of the curvature profile as~\cite{Young:2019yug}
\be
\label{threhsold_zeta}
\zeta_c = \frac{e}{2 \gamma} \lp 1- \sqrt{1-\frac{3}{2}\delta_c} \rp.
\ee
While we will use the parametrization \eqref{parametrization} of the curvature perturbation profile in Section 4,  we stress that, being the curvature perturbation a stochastic quantity, one may only calculate the mean profile of the curvature perturbations above a threshold. 

As we will show in the next section, both the probability of the smoothed density contrast and the knowledge of the curvature perturbation profile may be obtained employing a path-integral approach in terms of the $n$-point  correlators of the curvature perturbation. This requires however the knowledge of its joint probability.

\section{What is required to calculate the PBH abundance}
\renewcommand{\theequation}{3.\arabic{equation}}
\setcounter{equation}{0}
\noindent
In this section we are going to show, with the same path integral technique used in the literature to study the non-Gaussian fluctuations in the context of large scale structure~\cite{1986ApJ...310...19G,Matarrese:1986et,Bertschinger:1987qp, 1988PhRvL..61..267C}, the dependence of both the smoothed density contrast probability and the curvature perturbation profile in terms of the correlators of the curvature perturbation field.

\subsection{The smoothed density contrast probability}
\noindent
As stressed in the previous section, the key parameter to compute the PBH probability is the smoothed linear density contrast $\delta_l$. The joint probability that, at a given spatial point $\vec x$, the linear field $\delta_l(\vec x)$ attains a value in the range between 
$\delta_l$ and $\delta_l+\d \delta_l$ is given using the path integral approach as
\be
P(\delta_l)=\int [{\cal D}\zeta(\vec x)]\,P[\zeta(\vec x)]\,\delta_D[\delta_l(\vec x)-\delta_l],
\ee
where $\delta_D[\cdots]$ indicates the Dirac-delta distribution, while $P[\zeta(\vec x)]$ represents the probability distribution of the curvature field.
Given that
\be
\delta_l = -\frac{4}{3}r_m\zeta'(r_m) = -\frac{4}{9a^2H^2}\frac{3}{R^3}\int\,\d r \,r^2\,\theta(R-r)\nabla^2\zeta \bigg|_{R = r_m},
\ee
where $\theta$ indicates the Heaviside step function, we can restrict to the calculation of the more general integral
\be
\label{Eq.Palpha}
P(\alpha)=\int [{\cal D}\zeta(\vec x)]\,P[\zeta(\vec x)]\,\delta_D[\nabla^2\zeta_R(\vec x)-\alpha],
\ee
where we have introduced the quantity
\be
\nabla^2\zeta_R(\vec x) \equiv \int \d^3 y \, W_R(|\vec x-\vec y|)\nabla^2\zeta (\vec y) =\frac{3}{R^3}\int\,{\rm d}r \,r^2\,\theta(R-r)\nabla^2\zeta,
\ee
choosing a top-hat window function in real space $W_R(|\vec x-\vec y|) = \theta(R-|\vec x-\vec y|)/V_R$ with $V_R = 4 \pi R^3/3$ and $|\vec x-\vec y| = r$, and performing the angular integral assuming isotropy.
To compute the functional integration in Eq.~\eqref{Eq.Palpha} one can make use of the integral representation of the Dirac-delta function 
\be
\delta_D(x)=\int_{-\infty}^{\infty}\,
\frac{\d \lambda}{2\pi }\,e^{i\lambda x},
\ee
to write
\be
P(\alpha)=\int_{-\infty}^{\infty}\,
\frac{\d \lambda}{2\pi} \int [{\cal D}\zeta(\vec x)]\,P[\zeta(\vec x)]\,
{\rm exp}\left[-i\lambda\alpha +i\lambda\int\d^3 y W_R(|\vec x-\vec y|)\nabla^2\zeta (\vec y)
\right].
\ee
By using the definition of the partition function
\be
\label{partition}
Z[J]=\int [{\cal D}\zeta(\vec y)]\,P[\zeta(\vec y)]\,{\rm exp}\left[i\int \d^3 y J(\vec y)\zeta(\vec y)\right],
\ee
in terms of the arbitrary external source $J$, we therefore get
\be
P(\alpha)=\int_{-\infty}^{\infty}\,
\frac{\d \lambda}{2\pi}\,e^{-i\lambda\alpha}Z[\widetilde J],\qquad \widetilde J(\vec y)=\left.\lambda \nabla_x^2W_R(|\vec x-\vec y|)\right|_{\vec x=0}.
\ee
The probability of PBH formation $P(\alpha)$ can then be expressed introducing the connected correlation functions of the curvature perturbation 
\be
\xi^{(n)}(\vec x_1,\cdots, \vec x_n) = \int [{\cal D}\zeta(\vec x)]\,P[\zeta(\vec x)]\, \zeta(\vec x_1) \cdots  \zeta(\vec x_n),
\ee
by expanding the logarithm of the partition function as
\be
\ln Z[\widetilde J]=\sum_{n=2}\frac{i^n}{n!}\int \d^3 x_1\cdots \d^3 x_n \,
\xi^{(n)}(\vec x_1,\cdots, \vec x_n)\widetilde J(\vec x_1)\cdots \widetilde J(\vec x_n) = \sum_{n=2}\frac{(i\lambda)^n }{n!} \xi_n(R),
\ee
where 
\be
\label{xiR}
\xi_n(R)\equiv \int \left[ \prod_{a=1}^n \d^3 x_a \left. \nabla_x^2 W_R(|\vec x_a-\vec x|)\right|_{\vec x=0}\right]\xi^{(n)}(\vec x_1,\cdots, \vec x_n).
\ee
The probability then becomes
\be
P(\alpha)=\frac{1}{\sqrt{2\pi}} \, {\rm exp}\left[
\sum_{n=2}\frac{(-1)^n}{n!} \xi_n(R)\left(\frac{\partial}{\partial\alpha}\right)^n\right]{\rm exp}\left[-\alpha^2/2\right].
\ee
Having chosen a top-hat window function for which
\begin{align}
V_R \nabla_x^2 W_R(|\vec x_a-\vec x|)|_{\vec x=0} = - 2\frac{\delta_D (R-|\vec x_a|)}{|\vec x_a|} - \frac{\partial  \delta_D (R-|\vec x_a|)}{\partial |\vec x_a|},
\end{align}
Eq.~\eqref{xiR} can be rewritten as (labelling $|\vec x_a| = r_a$)
\begin{align}
\xi_n(R)
& = - \lp\frac{4 \pi}{V_R}\rp^n \int \left[ \prod_{a=1}^n \d r_a r_a^2 \left(\frac{2}{r_a}\delta_D (R-r_a)+ \frac{\partial  \delta_D (R-r_a)}{\partial r_a}\right)  \right] \int \frac{\d \Omega_1}{4 \pi} \dots  \frac{\d \Omega_n}{4 \pi} \xi^{(n)}(\vec x_1,\cdots, \vec x_n) \nonumber \\
& \equiv  - \lp\frac{4 \pi}{V_R}\rp^n \int \prod_{a=1}^n \d r_a r_a^2 \left(\frac{2}{r_a}\delta_D (R-r_a)+ \frac{\partial  \delta_D (R-r_a)}{\partial r_a}\right)  \xi^{(n)}_{\rm av}( r_1,\cdots, r_n).
\end{align}
One finally gets
\begin{align}
\xi_n(R)&= 
\lp \frac{3}{R}\rp^n \prod_{a=1}^n \frac{\partial}{\partial r_a} \xi^{(n)}_{\rm av}( r_1,\cdots, r_n) \bigg|_{r_a = R}.
\end{align}
From this final equation one can then appreciate how, in order to properly compute the abundance of PBH formation, one needs the knowledge of the $n$-point correlation functions of the curvature field, which can be only computed knowing  the joint probability of the field in different spatial points.

\subsection{The smoothed density contrast threshold}
\noindent
As we previously stressed, the knowledge of the  threshold  of  the smoothed density contrast passes through the calculation of the mean profile of the curvature perturbations which are above a given threshold (or peaks). In other words, the threshold is not an universal number, it depends on the perturbation profile. The goal of this section is to show that the calculation of the threshold demands the knowledge of, at least,  the two-point joint probability of the curvature perturbation.

We follow Ref.~\cite{Kehagias:2019eil} where such calculations are already present and we summarise them here for the sake of the reader. The starting point is the conditional probability to have at a distance $r$ from a threshold a given value $\zeta(r)$ 
\begin{eqnarray}
P(\zeta(r)|\zeta_0>\nu\sigma)=\frac{\Big<\delta_D(\zeta(\vec x)-\zeta(r))\theta(\zeta_0-\nu\sigma)\Big>}{\Big<\theta(\zeta_0-\nu\sigma)\Big>}.
\end{eqnarray}
Here $\sigma$ is the square root of the variance of the curvature perturbation and the threshold is indicated by $\nu\sigma$.
The mean   curvature perturbation we look for  corresponds to
\begin{align}
\label{sim}
\overline{\zeta}(r)&=\langle \zeta(r)|\zeta_0>\nu\sigma\rangle =
\int_{-\infty}^{\infty}{\rm d}\zeta(r) \,\zeta(r)
\frac{P(\zeta(r),\zeta_0>\nu\sigma)}{P(\zeta_0>\nu\sigma)}\nonumber\\
&= \frac{1}{P(\zeta_0>\nu\sigma)}\int_{-\infty}^{\infty}{\rm
d}\zeta(r) \,\zeta(r)
\Big<\delta_D(\zeta(\vec x)-\zeta(r))\theta(\zeta_0-\nu\sigma)\Big>\nonumber\\
&= \frac{1}{P(\zeta_0>\nu\sigma)}\int_{-\infty}^{\infty}{\rm
d}\zeta(r) \,\zeta(r) \int [D\zeta(\vec x)]P[\zeta(\vec x)]
\delta_D(\zeta(\vec x)-\zeta(r))\theta(\zeta_0-\nu\sigma)\nonumber \\
&=\frac{1}{P(\zeta_0>\nu\sigma)} \int [D\zeta(\vec x)]P[\zeta(\vec x)]
\, \zeta(\vec x) \,  \theta(\zeta_0-\nu\sigma)=\frac{\Big<\zeta(\vec x)\,
\theta(\zeta_0-\nu \sigma)\Big>}{\Big<\theta(\zeta_0-\nu\sigma)\Big>}.
\end{align}
Using the definition (\ref{partition}), we can  write
\begin{eqnarray}
\Big<\zeta(\vec x_1)\, \theta(\zeta(\vec x_2)-\nu \sigma)\Big>=(2\pi)^{-2}\sigma
\int_{-\infty}^\infty {\rm d} a_1\, a_1\int_{\nu}^\infty
{\rm d} a_2\int_{-\infty}^\infty{\rm d} \phi_1 \int_{-\infty}^\infty
{\rm d} \phi_2 e^{-i\sigma(\phi_1a_1+\phi_2 a_2)}Z[J], \label{dd}
\end{eqnarray}
with 
\begin{eqnarray}
J(\vec x)= \sum_{i=1}^2 J_i(\vec x_i, \vec x), \qquad J_i(\vec x_i, \vec x) = \phi_i \delta_D(\vec x_i-\vec x).
\end{eqnarray} 
Recalling  the standard expansion for  $\ln Z[J]$ 
\begin{align}
\ln Z[J]&= \sum_{n=2}^\infty\frac{i^n}{n!}\int \d^3 y_1\cdots \int \d^3 y_n \,
\sum_{i_1=1}^2\cdots \sum_{i_n=1}^2 J_{i_1}(\vec y_1,\vec x_1)\cdots J_{i_n}(\vec y_n,\vec x_n)
 \xi^{(n)}(\vec y_1,\cdots,\vec y_n) \nonumber \\
 & = \sum_{n=2}^\infty\frac{i^n}{n!} 
 \sum_{m=0}^n{ {n}\choose{m}} \phi_1^m \phi_2^{n-m}
\xi^{(n)}_{[m,n-m]},
\end{align}
where 
\begin{equation}
\xi^{(n)}_{[m,n-m]} =
\xi^{(n)}\big(
\underbrace{\vec x_1,\cdots,\vec x_1}_{m\text{-times}},\underbrace{\vec x_2,\cdots,\vec x_2}_{(n-m)\text{-times}}\big),
\end{equation}
we obtain
\begin{align}
\Big<\zeta(\vec x_1)\, \theta(\zeta(\vec x_2)-\nu \sigma)\Big>&=(2\pi)^{-2}\sigma
\int_{-\infty}^\infty {\rm d} a_1\, a_1\int_{\nu}^\infty
{\rm d} a_2\int_{-\infty}^\infty{\rm d} \phi_1 \int_{-\infty}^\infty
{\rm d} \phi_2 \nonumber \\
& \exp\left\{  \sum_{n=2}^\infty\frac{(-1)^n}{n!} 
 {\sum_{m=0}^n}'{ {n}\choose{m}} 
\xi^{(n)}_{[m,n-m]}\frac{\partial^m}{\partial a_1^m}
\frac{\partial^{n-m}}{\partial a_2^{n-m}}
\right\}\nonumber \\
& \exp{\left(-\frac{1}{2}\sigma^2(\phi_1^2+\phi_2^2)-
i\sigma(\phi_1a_1+\phi_2a_2)\right)}.
\end{align}
The  prime on the sum indicates that it has to be performed by omitting  the terms containing $\phi_1^2$ and $\phi_2^2$. 
Integrating over the variables $\phi_1$ and $\phi_2$ 
\begin{align}
\Big<\zeta(\vec x_1)\, \theta(\zeta(\vec x_2)-\nu \sigma)\Big>& = (2\pi)^{-1}\sigma
\int_{-\infty}^\infty {\rm d} a_1\, a_1\int_{\nu}^\infty
{\rm d} a_2  \nonumber \\
&\exp\left\{  \sum_{n=2}^\infty\frac{(-1)^n}{n!} 
 {\sum_{m=0}^n}'{ {n}\choose{m}} 
\frac{1}{\sigma^n}\xi^{(n)}_{[m,n-m]}\frac{\partial^m}{\partial a_1^m}
\frac{\partial^{n-m}}{\partial a_2^{n-m}}
\right\}\exp\left(-\frac{1}{2}(a_1^2+a_2^2)\right) , \label{ss0}
\end{align}
 we find
 \begin{eqnarray}
 \label{ss1}
\Big<\zeta(\vec x_1)\, \theta(\zeta(\vec x_2)-\nu \sigma)\Big>=
\frac{\sigma}{\sqrt{2\pi}}e^{-\frac{\nu^2}{2}}\left(\frac{\xi^{(2)}(\vec x_1,\vec x_2)}{\sigma^2}+
\frac{\nu}{2 \sigma^3} \xi^{(3)}(\vec x_1,\vec x_2,\vec x_2)+\frac{\nu^2-1}{6 \sigma^4}
\xi^{(4)}(\vec x_1,\vec x_2,\vec x_2,\vec x_2)+\cdots\right). \nonumber \\
\end{eqnarray}
The  connected piece of Eq.~\eqref{ss1} turns out to be  
\begin{eqnarray}
\Big<\zeta(\vec x_1)\, \theta(\zeta(\vec x_2)-\nu \sigma)\Big>
=\frac{\sigma}{\sqrt{2\pi}}e^{-\nu^2/2}\sum_{m=0}^\infty \frac{1}{2^{m/2}(m+1)!}
\frac{1}{\sigma^{(m+2)}}\xi^{(m+2)}(\vec x_1,\vec x_2,\cdots,\vec x_2)H_{m}\left(\frac{\nu}{\sqrt{2}}\right),\nonumber\\
&&
\end{eqnarray}
where $H_m(x)$ are the Hermite polynomials. At the same time, 
the  one-point non-Gaussian threshold probability   for $\nu\gg 1$  is 
\be
\Big<\theta(\zeta_0-\nu\sigma)\Big>\approx \frac{e^{-\nu^2/2}}{\sqrt{2\pi}\nu}{\rm exp}\left(\sum_{n=3}^\infty (\nu/\sigma)^n \xi^{(n)}(0)/n!\right),
\ee
where $\xi^{(n)}(0)$ are the $n$-point correlators calculated at the same point.  The  mean profile at distance $r$ from the 
origin and for large thresholds is given by ($|\vec x_2-\vec x_1|=r$) 
\begin{eqnarray}
\label{final}
\overline{\zeta}(r)=\nu\left[\frac{\xi^{(2)}(r)}{\sigma}+
\frac{\nu}{2\sigma^2} \xi^{(3)}(\vec x_1,\vec x_2,\vec x_2)+\frac{\nu^2}{6\sigma^3}
\xi^{(4)}(\vec x_1,\vec x_2,\vec x_2,\vec x_2)+\cdots\right]{\rm exp}\left(-\sum_{n=3}^\infty (\nu/\sigma)^n \xi^{(n)}(0)/n!\right).\nonumber\\
&&
\end{eqnarray}
This expression, which is the starting point for the calculation of the
smoothed density contrast threshold, shows that the knowledge of the PBH abundance asks at least for the bivariate joint probability of the curvature perturbation $P[\zeta(\vec x_1),\zeta(\vec x_2)]$, necessary to compute all the connected $n$-point correlators evaluated in two different points. 

Another, maybe more intuitive,  way to understand why the correlators of the curvature perturbation are needed is the following. Take a peak of the curvature perturbation, located at $\vec x=0$ where the value of the curvature perturbation is $\zeta_0$, which by itself is a stochastic variable. Performing a rotation of the coordinate axes to be aligned with the principal axes of length
$\lambda_i$ ($i=1,2,3$) of the constant-curvature perturbation ellipsoid and Taylor expanding up to second-order gives~\cite{Bardeen:1985tr} 
\be
\zeta(r)\simeq \zeta_0-\frac{1}{2}\sum_{i=1}^3\lambda_ix_i^2, \qquad r^2=\sum_{i=1}^3x_i^2.
\ee
It is straightforward to obtain 
\be
\delta_l= -\frac{4}{3}r_m\zeta'(r_m)\simeq \frac{8}{3}\left[\zeta_0-\zeta(r_m)\right].
\ee
This expression shows that  the statistics of the 
smoothed density contrast calculated in a volume of radius $r_m$ demands the knowledge of the correlations of the curvature perturbation in two different spatial points.

\section{A simplified comparison}
\renewcommand{\theequation}{4.\arabic{equation}}
\setcounter{equation}{0}
\noindent
Given the result of the previous section, one is obliged to ask how far we can go in the calculation of the PBH abundance. Finding that the tail of the probability of the curvature perturbation is not  Gaussian but, e.g. an exponential function, is not enough. On the contrary, it makes the story more intricate. Indeed, while for a Gaussian variable the joint probability is a multivariate normal distribution, for other functions the joint probabilities are not unique. 

One option is, of course, to know the expression of the non-linear curvature perturbation in terms of a Gaussian component. If so, one may and must  calculate the joint probability of the curvature perturbation from the  multivariate normal distribution of the linear Gaussian component\footnote{Notice also that the comparison between the abundance of PBHs calculated from the fully non-linear $\zeta$ and the one computed with its Gaussian component adopting the same threshold, as routinely done in the literature,  is in fact meaningless as the corresponding thresholds are different. They indeed depend on the peak profiles which are different in the two cases.}. In this section we are less ambitious, and we offer one example to  show that the abundance calculated from the curvature perturbation is different from the one inferred from the smoothed density contrast.  We do so by starting from the benchmark profile (\ref{parametrization}).

The example is the same used in Ref.~\cite{Biagetti:2021eep}, in which the authors considered
an ultra-slow-roll scenario of single-field models of inflation, which is based on the assumption that during its evolution, the inflaton field $\phi$ rolls down a flat region of its potential for a long enough period, during which violation of the slow-roll conditions results in an enhancement of the curvature perturbation. Using the $\delta N$ formalism, one can relate the perturbation of the inflaton field $\delta \phi$ to the curvature perturbation as
\be
\label{zeta}
\zeta=-\frac{1}{3}\ln\left(1+3\frac{\delta\phi}{\overline{\pi}_e}\right),
\ee
where $\overline{\pi}_e$ denotes the background velocity of the inflaton field at the end of the ultra-slow-roll phase.

Given the expected Gaussian behaviour of the massless perturbations $\delta \phi$ in de Sitter, one can determine the non-Gaussian probability distribution of the curvature perturbation by using probability conservation as
\begin{align}
\label{Pzeta}
P(\zeta)=P[\delta\phi(\zeta)]\left|\frac{{\rm d}\delta\phi}{{\rm d} \zeta}\right| = \frac{\overline{\pi}_e}{\sqrt{2 \pi} \sigma_{\delta \phi}} \exp \llp - \frac{1}{2 \sigma_{\delta \phi}^2} \lp \frac{\overline{\pi}_e^2}{9} \lp e^{-3 \zeta} - 1 \rp^2 \rp - 3 \zeta \rrp,
\end{align}
in terms of the variance of the Gaussian inflaton fluctuations 
\begin{eqnarray}
\sigma_{\delta\phi}^2=\int {\rm d}\ln k \,{\cal P}_{\delta\phi}(k),
\end{eqnarray}
and the inflaton perturbation power spectrum ${\cal P}_{\delta\phi}(k)$. As one can easily appreciate, in the limit of small curvature fluctuations  one recovers a Gaussian distribution, while for large values the probability is exponential and has a  non-Gaussian exponential tail\footnote{This result confirms what predicted in Refs.~\cite{Ezquiaga:2019ftu, Figueroa:2020jkf, Pattison:2021oen} using a stochastic approach, for large field displacement during the ultra-slow roll phase, for rapid subsequent transition into the slow-roll phase~\cite{Cai:2018dkf} and for a constant potential.}.

Given the exact expression of the non-linear curvature perturbation in terms of the Gaussian inflaton fluctuations, one can estimate the probability distribution of the smoothed linear density contrast $\delta_l$ by using the relation $\delta_l = - 4 r_m \zeta'(r_m)/3$ and conservation of probability as~\cite{Biagetti:2021eep}
\begin{align}
\label{Pdeltal}
P(\delta_l)&=\frac{e^{-1/2\sigma_Y^2}}{2\pi(\sigma_X^2+\sigma_Y^2 \delta_l^2)^{3/2}}\sigma_X\cdot\left[2\sigma_Y\sqrt{\sigma_X^2+\sigma_Y^2 \delta_l^2} \right. \nonumber \\
& \left. + \sqrt{2\pi}\sigma_X e^{\sigma_X^2/(2\sigma_X^2\sigma_Y^2+2\sigma_Y^4 \delta_l^2)}{\rm Erf}\left(\frac{\sigma_X}{\sqrt{2}\sigma_Y\sqrt{\sigma_X^2+\sigma_Y^2 \delta_l^2}}\right)\right], 
\end{align}
in terms of the variances
\begin{eqnarray}
\sigma_X^2=\frac{16r_m^2}{9\overline{\pi}_e^2}\int {\rm d}\ln k \,k^2 \,{\cal P}_{\delta\phi}(k),
\
\sigma_Y^2=\frac{9}{\overline{\pi}_e^2} \sigma^2_{\delta\phi}.
\end{eqnarray}
The probability function recovers a Gaussian in the limit of small inflaton fluctuations, while tends to a Cauchy distribution for large values of the field.

One can finally compare the PBH formation probability depending on the use of the smoothed density contrast $\delta_l$ or the curvature perturbation $\zeta$ as
\be
\beta_{\delta_m} =\int_{\delta_{c}}^{2/3} P(\delta_m) \d \delta_m= \int_{\delta_{l,c}}^{4/3} P(\delta_l) \d \delta_l \qquad \text{vs} \qquad 
\beta_\zeta = \int_{\zeta_c}^{\infty} P(\zeta) \d \zeta,
\ee
using the results of Eq.~\eqref{Pdeltal} and Eq.~\eqref{Pzeta} for the probability distributions and Eq.~\eqref{threhsold_delta} and Eq.~\eqref{threhsold_zeta} for the threshold to collapse.
In other words, instead of taking the mean profile, we adopt here  for simplicity the benchmark profile for the curvature perturbation (\ref{parametrization}).

\begin{figure}[t!]
	\centering
	\includegraphics[width=0.65\columnwidth]{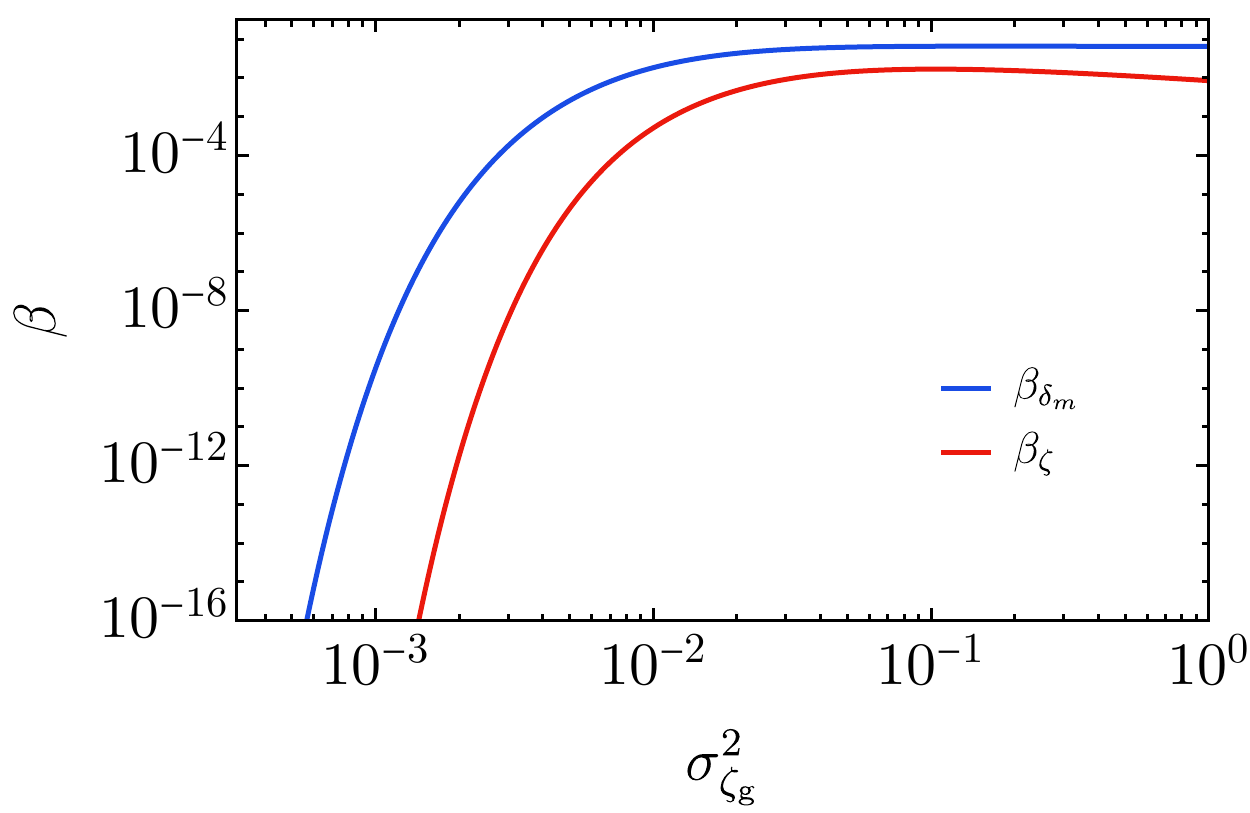}
	\caption{Comparison of the PBH abundance using the smoothed density field (blue line) and the curvature perturbation (red line), in terms of the rescaled variance of the Gaussian curvature perturbation component.}
	\label{figbeta}
\end{figure}

The comparison is shown in Fig.~\ref{figbeta} for the choice of a monochromatic power spectrum of the inflaton perturbations peaked at the characteristic momentum scale $k_*$, for which $\delta_c \simeq 0.59$ and  $k_* r_m \simeq 2.74$ have been computed numerically~\cite{Musco:2018rwt,Musco:2020jjb}. The corresponding value $\zeta_c \simeq 0.9$ is deduced  from Eq.~\eqref{threhsold_zeta}. We have also   introduced the normalised variance $\sigma_{\zeta_\text{\tiny g}} = \sigma_{\delta \phi}/\overline{\pi}_e$ of the Gaussian component.
As one can appreciate, the two approaches delivery a different abundance of PBHs, at least one order of magnitude different for large fluctuations.

\section{Conclusions}
\renewcommand{\theequation}{5.\arabic{equation}}
\setcounter{equation}{0}
\noindent
The goal of this short note was to clarify some confusion which seems to propagate in the literature about the  misuse of the curvature perturbation probability in the computation of the PBH abundance. Even though the non-Gaussian, non-perturbative, nature of a rare event like the formation of a PBH manifests itself in  the probability of the  curvature perturbation, we have stressed that the correct treatment should involve the smoothed density contrast and its threshold. This, in turns, requires 
the knowledge of all the $n$-point connected correlators of the curvature perturbation. Said in other words, one should in principle know the joint probabilities and not only the one-point probability $P(\zeta)$. 

\bigskip
\bigskip
\noindent
\subsection*{Acknowledgments}
\noindent
We thank M. Biagetti, G. Franciolini and A. Kehagias for useful discussions. 
V.DL. and  A.R. are supported by the Swiss National Science Foundation (SNSF), project {\sl The Non-Gaussian Universe and Cosmological Symmetries}, project number: 200020-178787.

\bibliographystyle{JHEP}
\bibliography{draft}

\end{document}